\documentclass[12pt]{article}
\usepackage[left=3cm,top=3cm,right=3cm,bottom=2cm]{geometry}
\usepackage{natbib}      
\usepackage{dcolumn}     
\usepackage{multirow}    
\usepackage{rotating}    
\usepackage{subeqn}
\usepackage{graphicx}
\usepackage{bm}
\author{Robert W. Johnson \\
\small Alphawave Research\\[-0.8ex]
\small Atlanta, GA, USA\\
\small \texttt{robjohnson@alphawaveresearch.com}\\}
\title{Influence of solar magnetic activity on the North American temperature record \\ {\large Comparison of the USHCN data with a wavelet analysis of the international sunspot number}}
\date{\today}
\renewcommand{\vec}[1]{\boldsymbol{#1}}

\newcommand{\beq}{\begin{equation}}
\newcommand{\eeq}{\end{equation}}
\newcommand{\bea}{\begin{eqnarray}}
\newcommand{\eea}{\end{eqnarray}}
\newcommand{\bes}{\begin{subeqnarray}}
\newcommand{\ees}{\end{subeqnarray}}
\newcommand{\ber}{\begin{sideways}}
\newcommand{\eer}{\end{sideways}}
\newcommand{\p}[2]{p^{#1}_{#2}}
\newcommand{\del}{\vec{\nabla}}

\begin{document}
%
%
%
%
%
%
%

\maketitle

\begin{abstract}
The effect of solar magnetic activity on the yearly mean average temperature is extracted from the historical record for much of North America.  The level of solar activity is derived from the international sunspot number by the renormalized continuous wavelet transform using the Morlet basis to provide a running estimate of the power associated with the magnetic cycle.  The solar activity gives the abscissa for a scatter plot of temperature for each station, from which the solar dependence and mean temperature are extracted.  These parameters are then plotted against the latitude, longitude, and elevation for each station, revealing a dependence of their values on geophysical location.  A mechanism to explain the latitudinal variation of the solar dependence is suggested.
%
\end{abstract}


\section{Introduction}
\label{intro}
The amount of influence which the solar magnetic cycle has on Earth's climate continues to be a matter of debate~\citep{lean-3069,2005ClDy25205W,moore-L17705,moore-34918,prsa:1452085,echera-41,li-1465111,2009ClDy321S}.  The intriguing coincidence of the Maunder Minimum~\citep{eddy06181976,Luterbacher2001}, a period of exceptionally low solar activity as indicated by a dearth of sunspots observed from 1645 and 1715, and the coldest years of the Little Ice Age~\citep{Grove2001,mann:02}, a period of exceptionally low temperatures as observed in Europe, North America, and elsewhere~\citep{KJK:08291997,TCJ:012001,hmst:9749} ending in the middle of the 19th century, suggests that the temperature on Earth might display a dependence on the level of solar magnetic activity.  Elucidating that dependence is this article's primary topic of investigation.

Using the renormalized continuous wavelet transform to provide a running estimate of the solar activity from the international sunspot number~\citep{rwj:jgr02,rwj:astro01}, here we extend its comparison to the yearly mean average temperature provided by the United States Historical Climatology Network, or USHCN~\citep{NDP:019,CDIAC:30,CDIAC:87}, for 1218 stations covering much of North America.  The mean average temperature series as given does not come with an error estimate (though one may be derived from the underlying monthly measurements and is the subject of further investigation), thus one unit of variance is assumed throughout.  From these data we extract the solar dependence and mean temperature for each station.  

Our basic strategy is to replace the abscissa of time in the temperature record for a station with the corresponding level of solar magnetic activity, thereby determining from the parameters of a linear regression that station's temperature's dependence on solar activity as well as the mean temperature without regard to the activity level.  Those parameters then become the ordinate against the three abscissas of station coordinates.  The distribution of these parameters displays a correlation with the geophysical location of the station, particularly for latitude and elevation but not so much for longitude, and another linear regression reveals their dependence.  The latitudinal variation of the solar dependence might be explained by an increased polar influx of energetic particles from the solar wind during periods of greater solar magnetic activity.

\section{Solar magnetic activity}
\label{sec:sma}
To evaluate the level of solar activity from the historical sunspot record, we use the renormalized continuous wavelet transform.  The algorithm is discussed in detail in~\citet{rwj:jpamt01}, thus only some highlights are given here.  Writing the Morlet wavelet at scale $s>0$ as the product of a real constant $C_s$, a Gaussian window $\Phi_s \equiv \exp (- \eta^2/2)$, and a normalized wave $\Theta_s \equiv \exp (i \omega_1 \eta)$ in terms of the parameter $\eta \equiv (t' - t) / s$, \bes
\psi_{s,t}(t') &\equiv& C_s \Phi_{s,t}(t') \Theta_{s,t}(t') \;, \\
 &=& \sqrt{2} \pi^{-1/4} s^{-3/2} e^{- \eta^2/2} e^{i \omega_1 \eta} \;, 
\ees where $\omega_1 \approx 2 \pi$ is the central frequency of the mother wavelet at unity scale $s=1$ and zero offset $t=0$ which here spans $2 \chi + 1 = 13$ time units $\Delta t \equiv 1$, the normalization $C_s = (4 / \pi s ^6)^{1/4}$ produces a wavelet with norm $\sqrt{2}/s$ and a symmetric forward and inverse transform pair given a mean-subtracted signal $y(t) \rightarrow y(t) - \langle y(t) \rangle$ of duration $N_t$, \bea
{\rm CWT}(s,t) &\equiv& \sum_{t'} \psi_{s,t}^*(t') y(t') \;,\\
 {\rm ICWT}(t) &\equiv& Re \left[\sum_s \sum_{t'} \psi_{s,t}^*(t') {\rm CWT}(s,t') \Delta s \right] \;, 
\eea {\it cf.} Equations (6) and (9) by~\citet{Frick:1997426}.  One compensates for the wavelet truncation caused by the finite signal duration by taking $\psi_{s,t} \leftarrow \psi_{s,t} \sqrt{2 / s^2 \vert \psi_{s,t} \vert^2}$ which preserves the wavelet norm.  The peak normalized power spectral density ${\rm PSD}(s,t) \equiv \vert \sqrt{2} {\rm CWT} \vert^2$ is twice the square of the amplitude of the CWT and displays in the instant wavelet power ${\rm IWP}_t(s) \equiv {\rm PSD}(s,t)$ response peaks whose integrated area equals the sum of the squared amplitudes for an infinite signal with sinusoidal components of stationary amplitude and period.  The integrated instant power ${\rm IIP}(t) \equiv \sum_s {\rm PSD}(s,t) \Delta s$ then gives a running estimate of the instant signal power, and the mean wavelet power ${\rm MWP}(s) \equiv N_t^{-1} \sum_t {\rm PSD}(s,t)$ provides the net power spectrum.  We have found satisfactory if not superior performance for data analysis when neglecting the admissibility condition~\citep{Frick:1997426,rwj:astro01} related to the DC response (nonzero mean) of the analyzing wavelet $\sum_{t'} \psi_{s,t}(t') \neq 0$, which remains negligibly small until the wavelet truncation becomes significant.

\begin{figure}
\includegraphics[width=8.4cm]{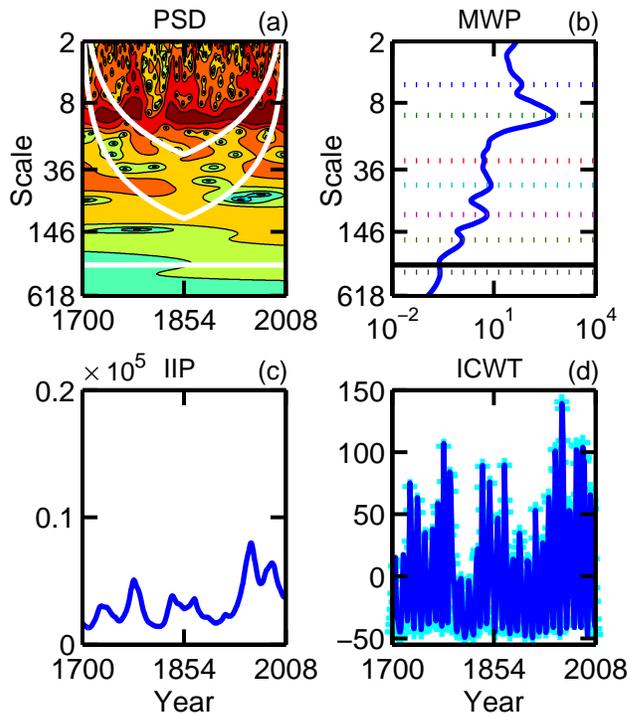}
\caption{CWT analysis of the international sunspot number whose IIP gives the spot number activity SNA}
\label{fig:A}       
\end{figure}

Since the invention of the telescope, solar astronomers have been able to keep an instrumental record of the magnetic flux erupting from the surface of the sun, which appears as darkened spots.  First formulated by Wolf, the international sunspot number $R_i$ by the~\citet{sidc:online} continues the Z\"{u}rich number up until the present day~\citep{2002JAVSO3148H} and is known to display a tighter correlation with sunspot area and radio flux~\citep{hathaway-357} than the sunspot group number $R_g$ by~\citet{hoytschatten:98}.  The CWT analysis of the mean-subtracted yearly smoothed sunspot number $R_i$ is presented in Figure~\ref{fig:A}, where the PSD has indicated in white the cone-of-influence, where wavelet compensation (which begins at the more restrictive cone-of-admissibility) becomes significant, as well as the scale of the signal duration at 309 years, beyond which lies the extremely low frequency ELF region.  Prominent peaks in the MWP are indicated by the dotted lines.  Its IIP gives the spot number activity SNA and outlines the magnitude of $R_i$ occuring at the maxima of the Schwabe cycle roughly every 11 years.

\begin{figure}
\includegraphics[width=8.4cm]{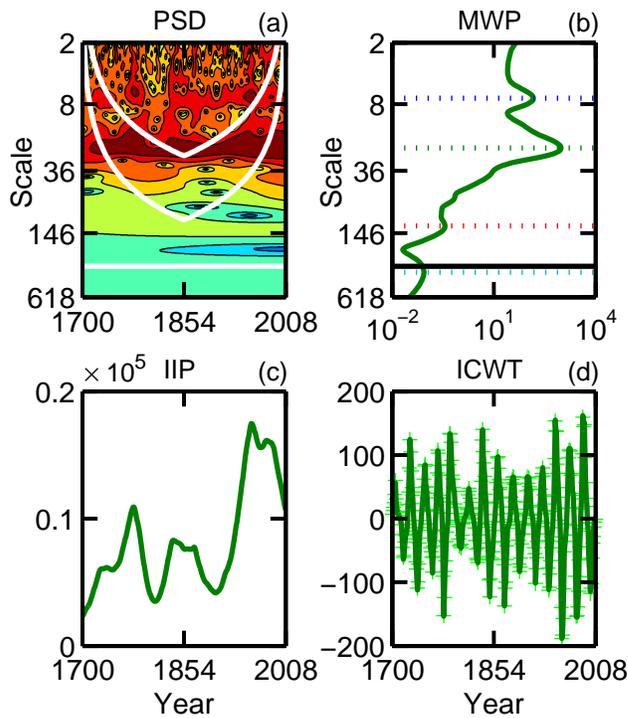}
\caption{CWT analysis of the derectified international sunspot number whose IIP gives the solar magnetic activity SMA}
\label{fig:B}       
\end{figure}

Underlying the sunspot number cycle is the solar magnetic activity discovered by~\citet{1908ApJ28315H} with a period of about 22 years and extending over the entire solar surface~\citep{1961ApJ133572B}.  The magnetic field of the spots reverses polarity over one Hale cycle, thus the sunspot number represents in some sense a rectified version of the solar magnetic activity signal.  In accord with the investigation by~\citet{BuckandMac:1992,BuckandMac:1993} and after a thorough evaluation of the arguments by~\citet{bracewell-53,bracewell-88}, we apply alternating signs to each of the Schwabe cycles in $R_i$ to form the derectified solar magnetic signal $\pm R_i$.  Its CWT analysis, shown in Figure~\ref{fig:B}, displays a much simpler spectral content dominated by the Hale cycle and its third harmonic at about 7 years.  Its IIP gives the solar magnetic activity SMA and is approximately double the SNA, with either denoted by SA.  Both the SNA and the SMA compare well with the power estimate found by~\citet{cjaa:46578} determined from just the Schwabe cycle.  While aware of the work by~\citet{1538:L154}, we maintain the traditional numbering of a tenth Schwabe cycle beginning about 1800.

\section{USHCN temperature record}
\label{sec:hcn}
\begin{figure}
\includegraphics[width=8.4cm]{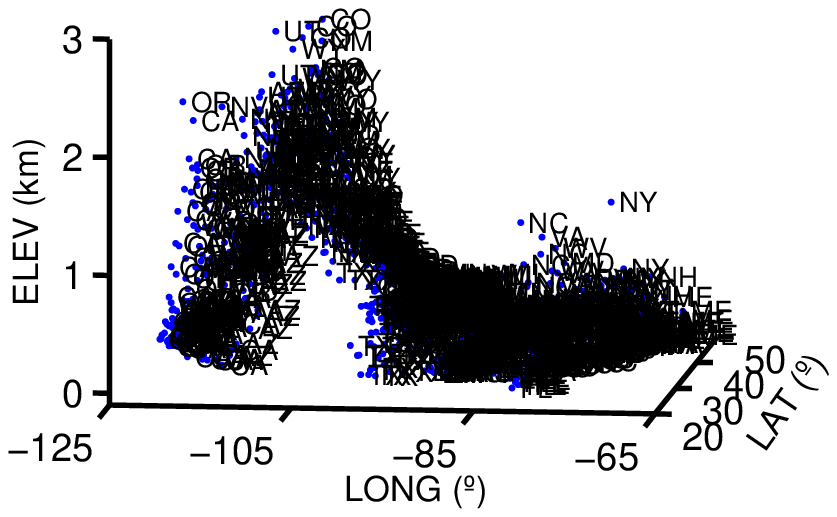}
\caption{Geophysical location of the USHCN stations}
\label{fig:C}       
\end{figure}

Unfortunately, the North American historical climate record does not begin shortly after the invention of the thermometer but gradually picks up contributing stations throughout the 19th and early 20th centuries.  For this analysis, we utilize the USHCN yearly mean average temperature data~\citep{NDP:019,CDIAC:30,CDIAC:87} as our climate signal.  The USHCN stations cover much of North America, as depicted in Figure~\ref{fig:C}, spanning approximately 30 degrees of latitude, 60 degrees of longitude, and 3 kilometers of elevation.  The density and extent of the station data reflects the duration of Western occupation as well as the accessibility of the location.  The values carry a flag indicating where missing data has been replaced with an estimate from surrounding data, and we will consider both the full data set as well as a restricted set consisting only of values flagged ``good''.

In performing our linear fits, we adopt a prior which is uniform on the angle of the slope rather than its magnitude, summarizing our methodology for those unfamiliar with Bayesian data analysis expressed in terms of conditional probabilities~\citep{Durrett-1994,Sivia-1996}.  Using notation $p(M \vert_I D) \equiv {\rm prob}(M \vert D, I)$ reading ``probability of $M$ given $D$ and information $I$'', we further abbreviate $p(M \vert_I D)$ to $p(M \vert D) \equiv \p{M}{D}$ and $p(M \vert I) \equiv \p{M}{}$ when the background information $I$ is unchanging.  For model $M$ with parameters $\vec{m} \equiv \{m_j\}$ and data $\vec{D} \equiv \{D_k\}$, Bayes' Theorem allows one to write $\p{M}{\vec{D}} = \p{\vec{D}}{M} \p{M}{} / \p{\vec{D}}{}$ reading ``the posterior for $M \vert \vec{D}$ equals the product of the likelihood for $\vec{D} \vert M$ and the prior for $M$ divided by the evidence for $\vec{D}$'', where the denominator $\p{\vec{D}}{}$ affecting neither parameter estimation nor model selection is often omitted.  The best estimate for the model parameters ${\vec{m}}_P$ is given by the maximum of the posterior $\p{\vec{m}}{\vec{D}} \propto \p{\vec{D}}{\vec{m}} \p{\vec{m}}{}$, whose logarithm (base $e$) is written as $L_P = L_L + L_{\vec{m}} - L_{\vec{D}}$, where the log evidence $L_{\vec{D}} = \#_{\vec{D}}$ is a constant.  

For a uniform prior $L_{\vec{m}} = \#_{\vec{m}}$, the maximum of the posterior is at the maximum of the likelihood ${\vec{m}}_P = {\vec{m}}_L$, and for independent data $\p{D_k}{D_l, \vec{m}} = \p{D_k}{\vec{m}}$ with Gaussian noise $\{\sigma_k\}$, \beq
\p{\vec{D}}{\vec{m}} = \prod_k \p{D_k}{\vec{m}} =  \prod_k (2 \pi \sigma_k^2)^{-1/2} \exp (-R_k^2/2) \;,
\eeq where $R_k = (M_k - D_k) / \sigma_k$ is the weighted residual for the $k$th datum, so that ${\vec{m}}_L$ minimizes the least-squares residual $\chi^2 \equiv \sum_k R_k^2$.  For a linear model $M_k \equiv y_k = m x_k + c$ giving the ordinate over an abscissa $\vec{x}$, the solution $\{m,c\}_L$ is found by setting $-\del L_L = \del \chi^2/2 = 0$, where \beq
\frac{1}{2} \del \chi^2 = \left[ \begin{array}{cc} S_{20} & S_{10} \\ S_{10} & S_{00} \end{array} \right] \left[ \begin{array}{c} m \\ c \end{array} \right] - \left[ \begin{array}{c} S_{11} \\ S_{01} \end{array} \right]
\eeq in terms of the weighted sums $S_{ab} \equiv \sum_k x_k^a D_k^b / \sigma_k^2$, and if one takes $x_k \rightarrow x'_k \equiv x_k - \langle \vec{x} \rangle$, the parameter $c_L$ returns the best estimate for the mean $\langle \vec{D} \rangle$ and the covariance matrix will be diagonal so that the error correlation $r_\sigma \equiv \sigma_{mc}^2 / (\sigma_{mm}^2 \sigma_{cc}^2)^{1/2}$ is zero, where the mean is computed with respect to weight $\sum_k \sigma_k^{-2}$.  Linearity implies that the shape of the posterior is a Gaussian given solely by the second derivative, \beq
- \del \del L_L = \frac{1}{2} \del \del \chi^2 = \left[ \begin{array}{cc} S_{20} & S_{10} \\ S_{10} & S_{00} \end{array} \right] = \left[ \begin{array}{cc} \sigma_{mm}^2 & \sigma_{mc}^2 \\ \sigma_{mc}^2 & \sigma_{cc}^2 \end{array} \right]^{-1} \;, 
\eeq whose width determines the covariance matrix.  With unit variance on the data, the deviation of the parameters depends only upon the values of the abscissa, which we assume are known exactly~\citep{Press-1992,Sivia-1996}.

A prior which is uniform on the magnitude of the slope $m$ will not be uniform on the angle of the slope $\theta$.  To show no preference for the slope of a line, one should take a prior which is uniform on its angle, and given unit bearing axes $\vec{x}$ and $\vec{y}$, that angle should be determined on the scale invariant axes $\vec{X}$ and $\vec{Y}$ to be defined.  First, one takes $\vec{X} \equiv \vec{x}' / \Delta_X$ for $\Delta_X \equiv {\rm max}(\vert x'_k \vert)$ to scale the abscissa to $[-1,1]$, and then the ordinate $\vec{Y} \equiv \vec{y} / \Delta_Y$ for $\Delta_Y \equiv {\rm max}(\vert D_k - \langle \vec{D} \rangle \vert)$ is not centered but is scaled as if it were.  On these axes, the line has slope $m' = m \Delta_X / \Delta_Y$ with angle $\theta = \arctan m'$, and it is on $\theta \in (-\pi/2, \pi/2)$ that we apply the uniform prior $\p{\theta}{} = 1/\pi$ to encompass slopes within the range $m \sim m' \in (-\infty,\infty)$.  In terms of $m'$, one writes $\p{\theta}{} = \p{m'}{} \vert {\rm d}m' / {\rm d}\theta \vert$ to find $\p{m'}{} = [\pi (1 + m'^2)]^{-1}$ such that $-L_{m'} = \log (1 + m'^2) + \#_\pi$, and then in terms of $m$ the prior is Lorentzian, \beq
\p{m}{} = \left \lbrace \frac{\Delta_Y}{\Delta_X} \pi \left[1 + \left( \frac{\Delta_X}{\Delta_Y} m \right)^2 \right] \right \rbrace^{-1} \;,
\eeq with normalization $\int \p{m'}{} {\rm d}m' = \int \p{m}{} {\rm d}m = 1$.  For comparison, a prior uniform on $m'$ will appear in $\theta$ as $\p{\theta}{} \propto 1 + (\tan \theta)^2$, which clearly displays a preference for a slope of extreme magnitude.

The prior nonuniform on $m \sim m'$ introduces a nonlinearity to the posterior, which becomes $L_P = L_L + L_m + L_c - L_{\vec{D}} = L_L + L_m + \#_{c, \vec{D}}$, as $\p{c}{} = 1/\Delta_c$ over a range of $\Delta_c$.  To find $m'_P$, one must solve $\del L_P = 0$, where \bes
-\del L_P &=& \frac{1}{2} \del \chi^2 + \del \log (1+m'^2) \\
 &=& \left[ \begin{array}{cc} S'_{20} & S'_{10} \\ S'_{10} & S'_{00} \end{array} \right] \left[ \begin{array}{c} m' \\ c' \end{array} \right] - \left[ \begin{array}{c} S'_{11} \\ S'_{01} \end{array} \right] + \left[ \begin{array}{c} 2 m' / (1 + m'^2) \\ 0 \end{array} \right] \;,
\ees where $S'_{ab}$ are the weighted sums and $\{m',c'\}$ are the model parameters on the scale invariant axes $\vec{X}$ and $\vec{Y}$.  The equation for $c'$ remains linear, leaving a cubic equation to solve for $m'$.  The appropriate real root is found between $m'_L$ and 0, and the solution is then scaled $\{m',c'\}_P \rightarrow \{m,c\}_P$, as is the covariance matrix $- [\del \del L_P]^{-1}$, which differs from $- [\del \del L_L]^{-1}$ only by $S'_{20} \rightarrow S'_{20} + 2 (1 - m'^2) / (1 + m'^2)^2$.  The nonlinear term introduces higher order corrections to the shape of the posterior, which remains very close to a Gaussian unless dominated by the prior.

\begin{figure}
\includegraphics[width=8.4cm]{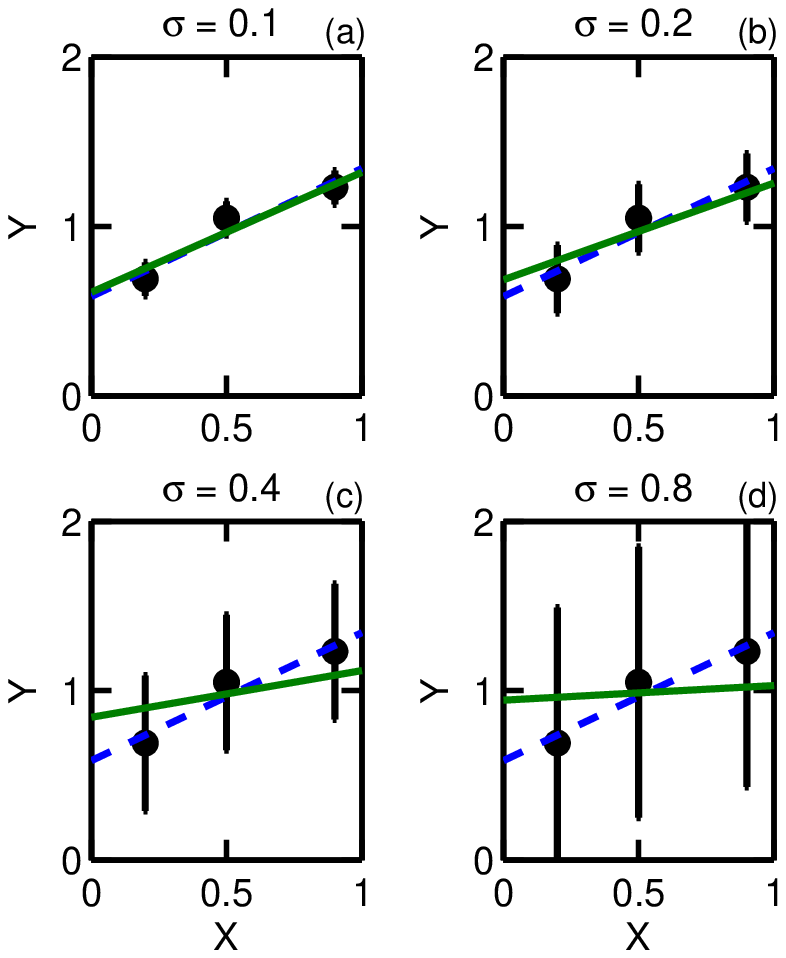}
\caption{Effect of the Lorentzian prior (solid) compared to the maximum likelihood estimate (dashed) as the variance increases}
\label{fig:D}       
\end{figure}

The net effect of the Lorentzian prior is to prevent one from overestimating the magnitude of the slope best supported by the evidence.  The estimate for the mean remains unaffected.  In the limit of poor data $\vert \del L_m \vert \gg \vert \del L_L \vert$, the best estimate for the slope $m_P$ is driven from $m_L$ towards zero.  We can see the effect of the prior on a fit to three points with increasing variance in Figure~\ref{fig:D}, where the maximum likelihood estimate is a dashed line.  One should consider how often one sees maximum likelihood used for cases such as (c) and (d), where the latter exhibits insufficient evidence for a significant slope.

\begin{figure}
\includegraphics[width=8.4cm]{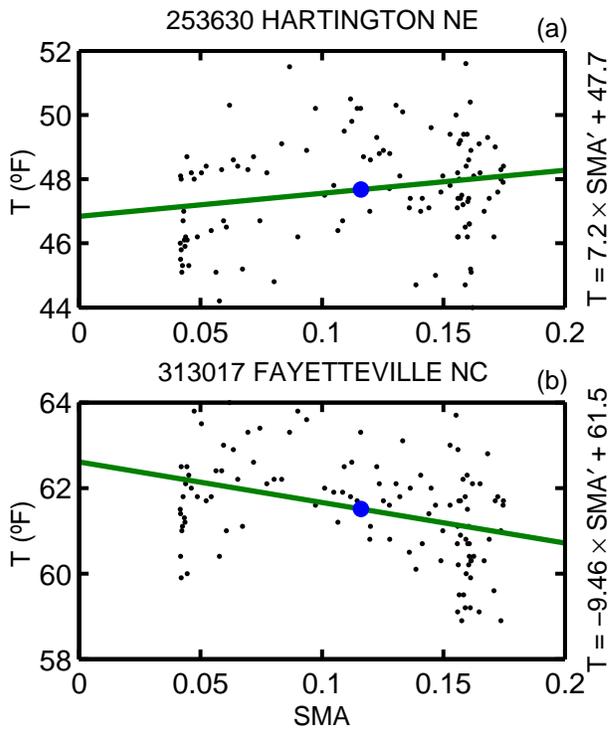}
\caption{Linear regression of mean average temperature $T$ against solar magnetic activity SMA using the full data set for two stations chosen at random}
\label{fig:E}       
\end{figure}

After so much digression, let us look at some data.  For each of the 1218 stations indexed by $j$ we take its temperature record and replace the abscissa of time with the value for the solar activity of that year (normalized by $10^5$). In Figure~\ref{fig:E} we display the linear regression for two locations chosen at random using the SMA and the full data set, and in Table~\ref{tab:A} we give the corresponding parameters for all four combinations of SA and data set.  The mean station temperature $T_0$, indicated by the heavy dot in the figure, is also the value of the line \beq
T^j({\rm SA}) = \Upsilon_{\rm SA}^j \times {\rm SA'} + T_0^j
\eeq at the mean of ${\rm SA}$ and indicates the temperature expected at that location without regard to solar magnetic activity.  The solar dependence $\Upsilon$ gives the correlation of the mean average temperature with the solar magnetic activity and for any particular station may be in either direction, as seen in Figure~\ref{fig:E}.

\begin{figure}
\includegraphics[width=8.4cm]{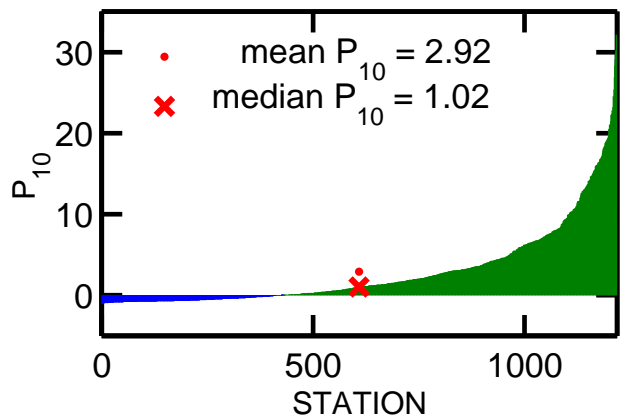}
\caption{Logarithm (base 10) of the relative probability of the linear model for all stations using the solar magnetic activity SMA and full data set with its mean and median values marked}
\label{fig:F}       
\end{figure}

There remains to answer the question of how much evidence there is for a nonvanishing solar dependence $\Upsilon$.  One cannot simply ask how good is the fit, as the only logical answer is, compared to what?  Bayesian data analysis frames the question in terms of model selection using the ratio of posteriors $P \equiv \p{B}{\vec{D}} / \p{A}{\vec{D}} = \p{\vec{D}}{B}\p{B}{} / \p{\vec{D}}{A}\p{A}{} = (\p{B}{} / \p{A}{}) (\p{\vec{D}}{B} / \p{\vec{D}}{A})$, where here model A has the single parameter $T_0$ and model B has two parameters $T_0$ and $\Upsilon$.  With no prior preference for either model $\p{B}{} / \p{A}{} = 1$, that ratio becomes \beq
P \rightarrow \frac{\p{\vec{D}}{B}}{\p{\vec{D}}{A}} = \frac{\int \int \p{\Upsilon}{B} \p{\vec{D}}{T_0,\Upsilon,B} {\rm d}T_0{\rm d}\Upsilon}{\int \p{\vec{D}}{T_0,A} {\rm d}T_0} \;,
\eeq where a common factor of $\p{T_0}{} = 1 / \Delta_{T_0}$ has been cancelled, whose logarithm (base 10) we write $P_{10}$.  For the remaining prior factor we restrict the range of the angle to $\pi/4$ so that $\p{\Upsilon}{B} = 4 \p{m}{}(m \rightarrow \Upsilon)$ by arguing that any child could draw a line through a given set of dots with a slope correct to within a slice of $\pi$, and in Table~\ref{tab:B} we give heuristic descriptions of several prior ranges which could be assigned.  The question essentially boils down to determining how much to penalize model B for having an extra parameter, and we feel that half a quadrant is sufficient, as without the prior factor model B would always be preferred.  In the numerator we take the quadratic approximation (which is exact for model A), $\p{\vec{D}}{B} \simeq 2 \pi \sigma_{\Upsilon \Upsilon} \sigma_{T_0 T_0}  \exp (-\chi^2_{T_0,\Upsilon}/2) \p{\Upsilon}{B}$ evaluated at the posterior maximum $\{T_0, \Upsilon\}_P$, after verifying its accuracy numerically.  In Figure~\ref{fig:F} we display $P_{10}$ for all stations using the SMA and the full data set, and in Table~\ref{tab:C} we give its mean and median for both SA and data sets.  We admit a preference for the SMA on the basis of the polarity associated with the Hale cycle, and we state subjectively that there is sufficient evidence to ascribe physical significance to the solar dependence $\Upsilon$.

\section{Results}
\label{sec:reslt}
\begin{figure}
\includegraphics[width=8.4cm]{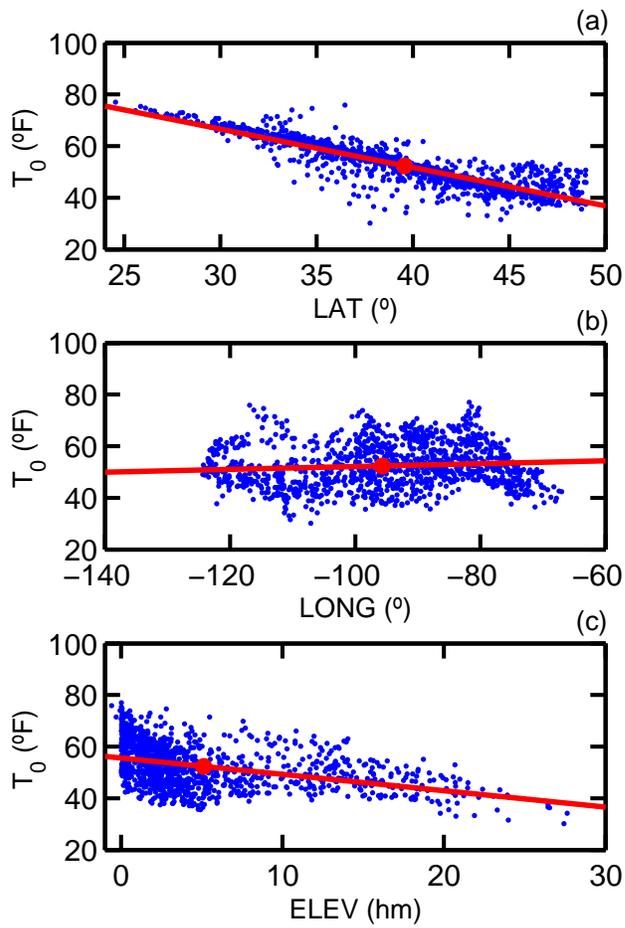}
\caption{Linear regression for the mean station temperature $T_0$ against geophysical location for the full data set}
\label{fig:G}       
\end{figure}

Collecting our parameters into two sets $\{T_0^j\}$ and $\{\Upsilon^j\}$, we proceed by plotting the ordered pairs $(T_0^j, X_i^j)$ and $(\Upsilon^j, X_i^j)$ for $i \in \{1,2,3\}$ and $\vec{X}^j \equiv ({\rm LAT}^j, {\rm LONG}^j, {\rm ELEV}^j)$ and perform another linear regression.  With so many points contributing, the prior here makes not much difference, but we maintain its use in our evaluation.  Beginning with the mean station temperature $T_0$, we plot its dependence against station location $\vec{X}^j$ in Figure~\ref{fig:G} for the SMA and full data set, and in Table~\ref{tab:D} are the results of the linear regression for all the combinations.  The column labelled $\sqrt{\langle \chi^2 \rangle}$ gives the rms weighted residual, {\it ie} the expected distance between the fit and a data point.  We see that $T_0$ displays the tightest correlation with latitude, as confirmed by the magnitude of Pearson's $r \equiv r_P$.  Its dependence on latitude and elevation is not unexpected, and its dependence on longitude is minor, most likely the result of correlation among the $\vec{X}^j$ representing the surface of the Earth.  The units (hm) of elevation are chosen so that its range is approximately equal to that of latitude, thus the relative dependence given by their slopes may be compared directly.  We see that the decrease in $T_0$ attributed to 1 degree of increasing latitude equates roughly to that attributed to 2 hm of increasing elevation and that the magnitudes of their $r_P$ are in about the same ratio.  The choice of solar activity signal does not affect $T_0$, and the choice of full or good temperature data has little impact on the results.

\begin{figure}
\includegraphics[width=8.4cm]{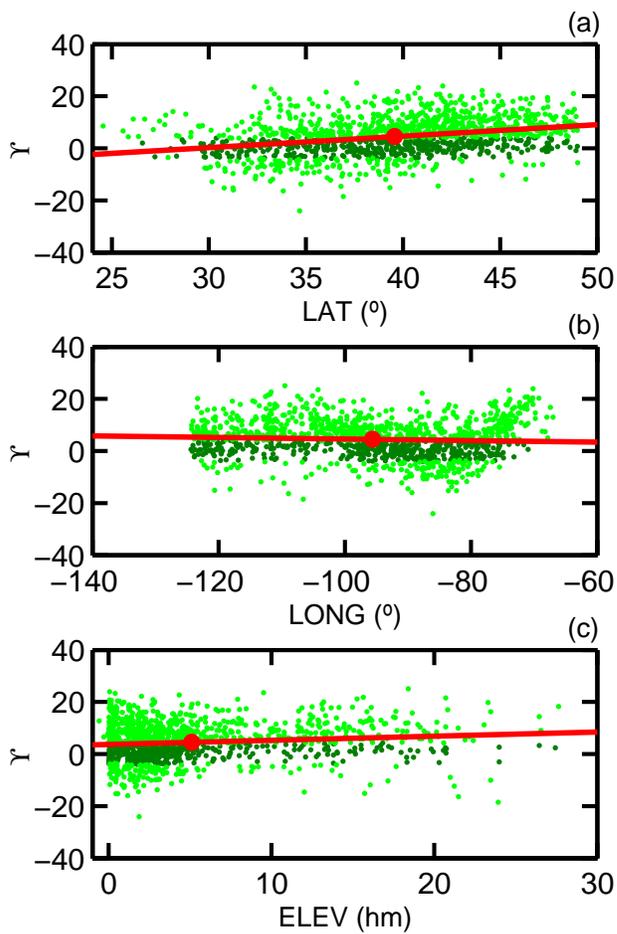}
\caption{Linear regression for the mean solar dependence $\Upsilon$ against geophysical location for the solar magnetic activity SMA and full data set}
\label{fig:H}       
\end{figure}

The solar dependence $\Upsilon$ also displays a dependence on location, which again is strongest for latitude.  Figure~\ref{fig:H} displays the SMA and full data set results, and Table~\ref{tab:E} gives the entire set of results.  The station averaged $\Upsilon$ is consistently positive, and we note a paucity of negative values at the lowest latitudes surveyed.  Dots for stations with little evidence for a slope are shaded darker and confined to a band around $\Upsilon = 0$.  The solar magnetic activity SMA results display a tighter correlation with location than those of the spot number activity SNA, and its units were chosen to make the visible range of $\Upsilon$ roughly the same as that of $T_0$ so that their relative slopes may also be compared.  In these units, the dependence of $\Upsilon$ on latitude is about a third the magnitude and opposite in direction to that for $T_0$, and its $r_P$ has been reduced also by about a third.  Its dependence on elevation is a third of that for latitude, and its variation with longitude is negligible.  The tighter correlation and smaller errors and residual indicate that the solar magnetic activity derived from the derectified sunspot number provides a more physical signal than the activity derived from the usual (rectified) sunspot number.  The choice of temperature data set does impact the results, with the full data set indicating a greater solar dependence.

\section{Discussion}
\label{discus}
Comparison of our work to others' is made difficult because so few investigators consider the derectified sunspot number in their analysis of solar magnetic activity and most employ a cross coherence analysis looking for cycles common to both the historical sunspot record and various indicators of climate~\citep{barnston-1295,zhaohan-42189,piscaron-1661,weber-917,barlyaeva-2009,2009ClDy321S}.  Our methodology here is quite different, eliciting a relationship between the net solar magnetic activity derived from the instant power carried by the derectified sunspot signal and the temperature recorded across much of North America for almost a century and a half.  What that data lacks in duration compared to the Central England Temperature record~\citep{rwj:jgr02,rwj:astro01} it more than makes up for in breadth of geophysical location.  We note that our results are in accord with other investigators~\citep{fcl-11011991,lean-3069,echera-41} and contrast with those who have found negligible evidence of solar forcing on Earth's climate~\citep{moore-L17705,moore-34918,prsa:1452085,li-1465111}.  Most authors, when speaking of the influence of solar forcing on Earth's climate, are actually discussing the effect of changing some model parameters in a {\it simulation} of meteorological systems~\citep{shind-4094S,2005ClDy25205W}, whereas here we are discussing a correlation found between independent instrumental observations from the historical record.

While several mechanisms have been proposed to explain the possible coupling between solar activity and terrestrial climate~\citep{svens-8122,lean-3069,hameedlee-32}, notably variation in irradiance, modulation of cosmic ray influx, and changes in global circulation, we would like to consider an alternative explanation.  From the dependence of $\Upsilon$ on latitude, we see that the degree of solar dependence increases with distance from the equator.  Thus, any proposed model for coupling should account for the greater dependence on solar magnetic activity towards the polar regions.  One effect of increased solar activity is an increased solar wind~\citep{webb94-4201}, whose temperature $10^5$K$\sim 10$eV is sufficient for ionization to the plasma state.  Those charged particles follow the lines of Earth's geomagnetic field emanating from the poles to produce the auroral illuminations and heating of the upper atmosphere.  By the well known conservation law, that energy must be deposited somewhere, and we propose that during times of increased solar activity, an increased influx of energy bearing solar wind particles produces a preferential heating towards the polar regions.  Whether that prediction will be supported by more detailed evaluations remains to be investigated.  The first step in such an evaluation would be to verify the relation between the energy flux from the solar wind and the solar magnetic activity, such as one can do for the 10.7cm radio flux~\citep{rwj-jaa01}, to put the historical sunspot record onto an axis bearing units relevant to geophysical analysis.  Then, with an estimate of the historical thermal flux variation, one could look for correlations with various climate indicators.

\section{Conclusions}
\label{concl}
A running estimate of the net solar magnetic activity is derived from the integrated instant power of the historical sunspot record, derectified to account for the opposite polarities of adjacent Schwabe cycles, using the renormalized continuous wavelet transform.  The net solar magnetic activity provides the abscissa for a linear regression using a Lorentzian prior of temperatures observed at stations covering much of North America for a duration exceeding one century.  A relationship is found between the level of solar activity and the observed temperature characterized by the slope of the linear regression which itself depends upon geophysical location.  The solar dependence is found to increase with latitude, suggesting a coupling mechanism to the polar regions of thermal flux from the solar wind.


\section*{Acknowledgements}
International sunspot number provided by the SIDC-team, World Data Center for the Sunspot Index, Royal Observatory of Belgium, Monthly Report on the International Sunspot Number, online catalogue of the sunspot index: http://www.sidc.be/sunspot-data/, 1700--2008.  Temperature data provided by the United States Historical Climatology Network, National Climatic Data Center, National Oceanic and Atmospheric Administration, http://cdiac.ornl.gov/epubs/ndp/ushcn/ushcn.html.  AlphaWavelet Toolbox Version 1.0 is available at http://www.alphawaveresearch.com.


\begin{thebibliography}{48}
\providecommand{\natexlab}[1]{#1}
\providecommand{\url}[1]{{#1}}
\providecommand{\urlprefix}{URL }
\expandafter\ifx\csname urlstyle\endcsname\relax
  \providecommand{\doi}[1]{DOI~\discretionary{}{}{}#1}\else
  \providecommand{\doi}{DOI~\discretionary{}{}{}\begingroup
  \urlstyle{rm}\Url}\fi
\providecommand{\eprint}[2][]{\url{#2}}

\bibitem[{{Babcock}(1961)}]{1961ApJ133572B}
{Babcock} HW (1961) {The Topology of the Sun's Magnetic Field and the 22-Year
  Cycle.} Astrophysical Journal 133:572--+, \doi{10.1086/147060}

\bibitem[{Barlyaeva et~al(2009)Barlyaeva, Mironova, and
  Ponyavin}]{barlyaeva-2009}
Barlyaeva T, Mironova I, Ponyavin D (2009) Nature of decadal variations in the
  climatic data of the second half of the 20th century. Doklady Earth Sciences
  425(2):419--423, \doi{10.1134/S1028334X09030155},
  \urlprefix\url{http://dx.doi.org/10.1134/S1028334X09030155}

\bibitem[{{Barnston} and {Livezey}(1989)}]{barnston-1295}
{Barnston} AG, {Livezey} RE (1989) A closer look at the effect of the 11-year
  solar cycle and the quasi-biennial oscillation on {N}orthern {H}emisphere 700
  mb height and extratropical {N}orth {A}merican surface temperature. J Climate
  2:1295–--1313

\bibitem[{{Bracewell}(1953)}]{bracewell-53}
{Bracewell} RN (1953) The sunspot number series. Nature 171:649--650

\bibitem[{{Bracewell}(1988)}]{bracewell-88}
{Bracewell} RN (1988) {Three-halves law in sunspot cycle shape}. Mon Not R Astr
  Soc 230:535--550

\bibitem[{Buck and Macaulay(1992)}]{BuckandMac:1992}
Buck B, Macaulay VA (1992) Entropy and sunspots: Their bearing on time-series.
  In: G E, P N, R SC (eds) Maximum Entropy and Bayesian Methods, Seattle 1991,
  Kluwer Academic Publishers, Netherlands, pp 241--252

\bibitem[{Buck and Macaulay(1993)}]{BuckandMac:1993}
Buck B, Macaulay VA (1993) Fine structure in the sunspot record. In:
  Mohammad-Djafari A, Demoments G (eds) Maximum Entropy and Bayesian Methods,
  Paris 1992, Kluwer Academic Publishers, Netherlands, pp 345--356

\bibitem[{{Durrett}(1994)}]{Durrett-1994}
{Durrett} R (1994) The Essentials of Probability. Duxbury Press, A Division of
  Wadsworth, Inc., Belmont, California, USA

\bibitem[{{Easterling} et~al(1996){Easterling}, Karl, Mason, Hughes, and
  Bowman}]{CDIAC:87}
{Easterling} DR, Karl TR, Mason EH, Hughes PY, Bowman DP (1996) {United States
  Historical Climatology Network (U.S. HCN) Monthly Temperature and
  Precipitation Data}. Tech. Rep. ORNL/CDIAC-87, NDP-019/R3, Carbon Dioxide
  Information Analysis Center, Oak Ridge National Laboratory, U.S. Department
  of Energy, Oak Ridge, Tennessee, USA

\bibitem[{Echer et~al(2009)Echer, Echer, Nordemann, and Rigozo}]{echera-41}
Echer MPS, Echer E, Nordemann DJR, Rigozo NR (2009) Multi-resolution analysis
  of global surface air temperature and solar activity relationship. Journal of
  Atmospheric and Solar-Terrestrial Physics 71(1):41--44,
  \doi{10.1016/j.jastp.2008.09.032},
  \urlprefix\url{http://dx.doi.org/10.1016/j.jastp.2008.09.032}

\bibitem[{Eddy(1976)}]{eddy06181976}
Eddy JA (1976) {The Maunder Minimum}. Science 192(4245):1189--1202,
  \doi{10.1126/science.192.4245.1189},
  \urlprefix\url{http://www.sciencemag.org/cgi/reprint/192/4245/1189.pdf}

\bibitem[{Frick et~al(1997)Frick, Baliunas, Galyagin, Sokoloff, and
  Soon}]{Frick:1997426}
Frick P, Baliunas S, Galyagin D, Sokoloff D, Soon W (1997) A wavelet analysis
  of solar chromospheric activity variations. ApJ 483:426--434,
  \eprint{http://www.journals.uchicago.edu/ApJ/journal/issues/ApJ/v483n1/34412%
/34412.web.pdf}

\bibitem[{{Friis-Christensen} and {Lassen}(1991)}]{fcl-11011991}
{Friis-Christensen} E, {Lassen} K (1991) {Length of the Solar Cycle: An
  Indicator of Solar Activity Closely Associated with Climate}. Science
  254(5032):698--700, \doi{10.1126/science.254.5032.698},
  \urlprefix\url{http://www.sciencemag.org/cgi/content/abstract/254/5032/698},
  \eprint{http://www.sciencemag.org/cgi/reprint/254/5032/698.pdf}

\bibitem[{Grove(2001)}]{Grove2001}
Grove JM (2001) The initiation of the {"Little Ice Age"} in regions round the
  north atlantic. Climatic Change 48(1):53--82, \doi{10.1023/A:1005662822136},
  \urlprefix\url{http://dx.doi.org/10.1023/A:1005662822136}

\bibitem[{{Hale}(1908)}]{1908ApJ28315H}
{Hale} GE (1908) {On the Probable Existence of a Magnetic Field in Sun-Spots}.
  Astrophysical Journal 28:315--+, \doi{10.1086/141602}

\bibitem[{Hameed and Lee(2005)}]{hameedlee-32}
Hameed S, Lee JN (2005) A mechanism for sun-climate connection. Geophys Res
  Lett 32:L23,817, \doi{10.1029/2005GL024393},
  \urlprefix\url{http://dx.doi.org/10.1029/2005GL024393}

\bibitem[{{Hathaway} et~al(2002){Hathaway}, {Wilson}, and
  {Reichmann}}]{hathaway-357}
{Hathaway} DH, {Wilson} RM, {Reichmann} EJ (2002) {Group Sunspot Numbers:
  Sunspot Cycle Characteristics}. Solar Physics 211:357--370

\bibitem[{Holmgren et~al(2001)Holmgren, Moberg, Svanered, and
  Tyson}]{hmst:9749}
Holmgren K, Moberg A, Svanered O, Tyson PD (2001) A preliminary 3000-year
  regional temperature reconstruction for {South} {Africa}. S Afr J Sci
  97:49--51

\bibitem[{{Hossfield}(2002)}]{2002JAVSO3148H}
{Hossfield} CH (2002) {A History of the Zurich and American Relative Sunspot
  Number Indices}. Journal of the American Association of Variable Star
  Observers (JAAVSO) 31:48--53

\bibitem[{{Hoyt} and {Schatten}(1998)}]{hoytschatten:98}
{Hoyt} DV, {Schatten} KH (1998) {Group Sunspot Numbers: A New Solar Activity
  Reconstruction}. Solar Physics 181:491--512

\bibitem[{{Johnson}(2009{\natexlab{a}})}]{rwj:jgr02}
{Johnson} RW (2009{\natexlab{a}}) {Enhanced wavelet analysis of solar magnetic
  activity with comparison to global temperature and the Central England
  Temperature record}. Journal of Geophysical Research (Space Physics)
  114(A05105), \doi{10.1029/2009JA014172},
  \urlprefix\url{http://www.agu.org/pubs/crossref/2009/2009JA014172.shtml},
  \eprint{arXiv:0812.2438}

\bibitem[{{Johnson}(2009{\natexlab{b}})}]{rwj:jpamt01}
{Johnson} RW (2009{\natexlab{b}}) Symmetrization and enhancement of the
  continuous {Morlet} transform. 
  \urlprefix\url{http://arxiv.org/abs/0912.1126}, under
  consideration, \eprint{arXiv:0912.1126}

\bibitem[{{Johnson}(2010{\natexlab{a}})}]{rwj:astro01}
{Johnson} RW (2010{\natexlab{a}}) {Edge adapted wavelets, solar magnetic
  activity, and climate change}. Astrophysics and Space Science pp 4--+,
  \doi{10.1007/s10509-009-0249-6},
  \urlprefix\url{http://www.springerlink.com/content/703288773149018m/},
  \eprint{arXiv:0911.4663}

\bibitem[{{Johnson}(2010{\natexlab{b}})}]{rwj-jaa01}
{Johnson} RW (2010{\natexlab{b}}) Power law relating 10.7 cm flux to sunspot
  number. 
  \urlprefix\url{http://arxiv.org/abs/0912.5042}, under consideration,
  \eprint{arXiv:0912.5042}

\bibitem[{Johnson et~al(2001)Johnson, Barry, Chan, and Wilkinson}]{TCJ:012001}
Johnson TC, Barry SL, Chan Y, Wilkinson P (2001) {Decadal record of climate
  variability spanning the past 700 yr in the Southern Tropics of East Africa}.
  Geology 29(1):83--86, \doi{10.1130/0091-7613(2001)029<0083:DROCVS>2.0.CO;2},
  \urlprefix\url{http://geology.gsapubs.org/content/29/1/83.abstract}

\bibitem[{{Karl} et~al(1990){Karl}, {Williams, Jr.}, and {Quinlan}}]{CDIAC:30}
{Karl} TR, {Williams, Jr} CN, {Quinlan} FT (1990) {United States Historical
  Climatology Network (HCN) Serial Temperature and Precipitation Data}. Tech.
  Rep. ORNL/CDIAC-30, NDP-019/R1, Carbon Dioxide Information Analysis Center,
  Oak Ridge National Laboratory, U.S. Department of Energy, Oak Ridge,
  Tennessee, USA

\bibitem[{Kreutz et~al(1997)Kreutz, Mayewski, Meeker, Twickler, Whitlow, and
  Pittalwala}]{KJK:08291997}
Kreutz KJ, Mayewski PA, Meeker LD, Twickler MS, Whitlow SI, Pittalwala II
  (1997) {Bipolar Changes in Atmospheric Circulation During the Little Ice
  Age}. Science 277(5330):1294--1296, \doi{10.1126/science.277.5330.1294},
  \urlprefix\url{http://www.sciencemag.org/cgi/content/abstract/277/5330/1294}

\bibitem[{Le(2004)}]{cjaa:46578}
Le GM (2004) Wavelet analysis of the {S}chwabe cycle properties in solar
  activity. Chin J Astron Astrophys 4(6):578--582,
  \urlprefix\url{http://www.chjaa.org/2004/2004\_4\_6p578.pdf}

\bibitem[{{Lean} and {Rind}(1998)}]{lean-3069}
{Lean} J, {Rind} D (1998) {Climate Forcing by Changing Solar Radiation}.
  Journal of Climate 11:3069--3094, \doi{10.1175/1520-0442(1998)011}

\bibitem[{Li et~al(2009)Li, Yang, Huang, and Li}]{li-1465111}
Li CH, Yang ZF, Huang GH, Li YP (2009) Identification of relationship between
  sunspots and natural runoff in the {Y}ellow {R}iver based on discrete wavelet
  analysis. Expert Syst Appl 36(2):3309--3318,
  \doi{http://dx.doi.org/10.1016/j.eswa.2008.01.083}

\bibitem[{Lockwood and Fr\"{o}hlich(2007)}]{prsa:1452085}
Lockwood M, Fr\"{o}hlich C (2007) Recent oppositely directed trends in solar
  climate forcings and the global mean surface air temperature. Proceedings of
  the Royal Society A: Mathematical, Physical and Engineering Sciences
  463:2447--2460, \doi{10.1098/rspa.2007.1880},
  \urlprefix\url{http://dx.doi.org/10.1098/rspa.2007.1880}

\bibitem[{Luterbacher et~al(2001)Luterbacher, Rickli, Xoplaki, Tinguely, Beck,
  Pfister, and Wanner}]{Luterbacher2001}
Luterbacher J, Rickli R, Xoplaki E, Tinguely C, Beck C, Pfister C, Wanner H
  (2001) The late {Maunder Minimum} (1675--1715) -- a key period for studying
  decadal scale climatic change in {Europe}. Climatic Change 49(4):441--462,
  \doi{10.1023/A:1010667524422},
  \urlprefix\url{http://dx.doi.org/10.1023/A:1010667524422}

\bibitem[{Mann(2002)}]{mann:02}
Mann ME (2002) Encyclopedia of Global Environmental Change, vol~1, John Wiley
  \& Sons, Ltd, Chichester, chap Little Ice Age, p 504–509

\bibitem[{Moore et~al(2006)Moore, Grinsted, and Jevrejeva}]{moore-L17705}
Moore J, Grinsted A, Jevrejeva S (2006) Is there evidence for sunspot forcing
  of climate at multi-year and decadal periods? Geophys Res Lett 33:L17,705,
  \doi{10.1029/2006GL026501},
  \urlprefix\url{http://dx.doi.org/10.1029/2006GL026501}

\bibitem[{Moore et~al(2007)Moore, Grinsted, and Jevrejeva}]{moore-34918}
Moore J, Grinsted A, Jevrejeva S (2007) Evidence from wavelet lag coherence for
  negligible solar forcing of climate at multi-year and decadal periods.
  Nonlinear Dynamics in Geosciences pp 457--464,
  \doi{10.1007/978-0-387-34918-3\_25},
  \urlprefix\url{http://www.springerlink.com/content/rt26l31g36262151}

\bibitem[{Piscaronoft et~al(2004)Piscaronoft, Kalvov\'{a}, and
  Br\'{a}zdil}]{piscaron-1661}
Piscaronoft P, Kalvov\'{a} J, Br\'{a}zdil R (2004) Cycles and trends in the
  {C}zech temperature series using wavelet transforms. International Journal of
  Climatology 24(13):1661--1670, \doi{10.1002/joc.1095},
  \urlprefix\url{http://dx.doi.org/10.1002/joc.1095}

\bibitem[{{Press} et~al(1992){Press}, {Teukolsky}, {Vetterling}, and
  {Flannery}}]{Press-1992}
{Press} WH, {Teukolsky} SA, {Vetterling} WT, {Flannery} BP (1992) Numerical
  Recipes. CUP, Cambridge, England

\bibitem[{{Quinlan} et~al(1987){Quinlan}, {Karl}, and {Williams,
  Jr.}}]{NDP:019}
{Quinlan} FT, {Karl} TR, {Williams, Jr} CN (1987) {United States Historical
  Climatology Network (HCN) Serial Temperature and Precipitation Data}. Tech.
  Rep. NDP-019, Carbon Dioxide Information Analysis Center, Oak Ridge National
  Laboratory, U.S. Department of Energy, Oak Ridge, Tennessee, USA

\bibitem[{{Shindell} et~al(2003){Shindell}, {Schmidt}, {Miller}, and
  {Mann}}]{shind-4094S}
{Shindell} DT, {Schmidt} GA, {Miller} RL, {Mann} ME (2003) {Volcanic and Solar
  Forcing of Climate Change during the Preindustrial Era.} Journal of Climate
  16:4094--4107, \doi{10.1175/1520-0442(2003)016<4094:VASFOC>2.0.CO;2}

\bibitem[{{SIDC-team}(1700-2008)}]{sidc:online}
{SIDC-team} (1700-2008) {The International Sunspot Number}. Monthly Report on
  the International Sunspot Number, online catalogue

\bibitem[{{Sitnov}(2009)}]{2009ClDy321S}
{Sitnov} SA (2009) {Influence of the 11-year solar cycle on the effects of the
  equatorial quasi-biennial oscillation, manifesting in the extratropical
  northern atmosphere}. Climate Dynamics 32:1--17,
  \doi{10.1007/s00382-007-0362-6}

\bibitem[{Sivia(1996)}]{Sivia-1996}
Sivia DS (1996) Data Analysis: a Bayesian Primer. OUP, Oxford, England

\bibitem[{Svensmark(1998)}]{svens-8122}
Svensmark H (1998) Influence of cosmic rays on earth's climate. Phys Rev Lett
  81(22):5027--5030, \doi{10.1103/PhysRevLett.81.5027}

\bibitem[{Usoskin et~al(2009)Usoskin, Mursula, Arlt, , and
  Kovaltsov}]{1538:L154}
Usoskin IG, Mursula K, Arlt R, , Kovaltsov GA (2009) A solar cycle lost in
  1793-1800: Early sunspot observations resolve the old mystery. The
  Astrophysical Journal Letters 700(2):L154--L157,
  \urlprefix\url{http://stacks.iop.org/1538-4357/700/L154}

\bibitem[{{Wagner} and {Zorita}(2005)}]{2005ClDy25205W}
{Wagner} S, {Zorita} E (2005) {The influence of volcanic, solar and
  $\mathrm{CO}_{2}$ forcing on the temperatures in the Dalton Minimum (1790
  1830): a model study}. Climate Dynamics 25:205--218,
  \doi{10.1007/s00382-005-0029-0}

\bibitem[{Webb and Howard(1994)}]{webb94-4201}
Webb DF, Howard RA (1994) The solar cycle variation of coronal mass ejections
  and the solar wind mass flux. J Geophys Res 99(A3):4201--4220,
  \doi{10.1029/93JA02742}, \urlprefix\url{http://dx.doi.org/10.1029/93JA02742}

\bibitem[{{Weber}(2005)}]{weber-917}
{Weber} SL (2005) {A timescale analysis of the Northern Hemisphere temperature
  response to volcanic and solar forcing}. Climate of the Past 1:9--17

\bibitem[{Zhao et~al(2004)Zhao, Han, and Li}]{zhaohan-42189}
Zhao J, Han YB, Li ZA (2004) The effect of solar activity on the annual
  precipitation in the beijing area. Chinese Journal of Astronomy and
  Astrophysics 4(2):189--197,
  \urlprefix\url{http://stacks.iop.org/1009-9271/4/189}

\end{thebibliography}

\newpage

\begin{table}
\centering
\caption{Linear regression results for two random stations}
\label{tab:A}       
\begin{tabular}{l|l|l|llllll}
\hline\noalign{\smallskip}
ID \# & data & SA & $\Upsilon$ & $T_0$ & $\sigma_{\Upsilon \Upsilon}$ & $\sigma_{T_0 T_0}$ & $P_{10}$  \\
\noalign{\smallskip}\hline\noalign{\smallskip}
\multirow{4}{*}{\ber 253630 \eer} & \multirow{2}{*}{\ber full \eer} & SMA & 7.2 & 47.7 & 2.01 & 0.0937 & 1.88 \\\cline{3-8}
 &  & SNA & 13.6 & 47.7 & 4.94 & 0.0937 & 0.826 \\\cline{2-8}
 & \multirow{2}{*}{\ber good \eer} & SMA & 5.32 & 47.7 & 2.12 & 0.0995 & 0.477 \\\cline{3-8}
 & & SNA & 9.52 & 47.7 & 5.22 & 0.0995 & -0.0718 \\
\noalign{\smallskip}\hline\noalign{\smallskip}
\multirow{4}{*}{\ber 313017 \eer} & \multirow{2}{*}{\ber full \eer} & SMA & -9.46 & 61.5 & 2.00 & 0.0937 & 4.11 \\\cline{3-8}
 &  & SNA & -20.6 & 61.5 & 4.93 & 0.0937 & 3.12 \\\cline{2-8}
 & \multirow{2}{*}{\ber good \eer} & SMA & -9.82 & 61.6 & 2.17 & 0.105 & 3.72 \\\cline{3-8}
 & & SNA & -20.8 & 61.6 & 5.31 & 0.105 & 2.72 \\
\noalign{\smallskip}\hline
\end{tabular}
\end{table}

\begin{table}
\centering
\caption{Heuristic description of various prior ranges}
\label{tab:B}       
\begin{tabular}{c|lr}
\hline\noalign{\smallskip}
prior range &  \multicolumn{2}{l}{the scattered dots are ...} \\
\noalign{\smallskip}\hline\noalign{\smallskip}
$\pi$   & & there \\
$\pi/2$ & & going up or going down \\
$\pi/3$ & & going up, holding steady, or going down \\
$\pi/4$ & & sharply up, slightly up, slightly down, or sharply down \\
\noalign{\smallskip}\hline
\end{tabular}
\end{table}

\begin{table}
\centering
\caption{Mean and median $P_{10}$}
\label{tab:C}       
\begin{tabular}{l|l|ll}
\hline\noalign{\smallskip}
data & SA & mean $P_{10}$ & median $P_{10}$  \\
\noalign{\smallskip}\hline\noalign{\smallskip}
\multirow{2}{*}{\ber full \eer} & SMA & 2.9172 & 1.0178 \\\cline{2-4}
 & SNA & 2.073 & 0.51658 \\\cline{1-4}
\multirow{2}{*}{\ber good \eer} & SMA & 1.6315 & 0.19813 \\\cline{2-4}
 & SNA & 1.0944 & -0.02401 \\
\noalign{\smallskip}\hline
\end{tabular}
\end{table}

\clearpage

\begin{table}
\centering
\caption{Linear regression results for $T_0$}
\label{tab:D}       
\begin{tabular}{l|l|llllll}
\hline\noalign{\smallskip}
$X_i$ & data & $m$ & $c$ & $\sigma_{mm}$ & $\sigma_{cc}$ & $\sqrt{\langle \chi^2 \rangle}$ & $r_P$ \\
\noalign{\smallskip}\hline\noalign{\smallskip}
\multirow{2}{*}{\ber LAT \eer} & full & -1.48 & 52.4 & 0.000539 & 0.00269 & 4.01 & -0.879 \\\cline{2-8}
 & good & -1.48 & 52.3 & 0.000636 & 0.00314 & 3.86 & -0.884 \\
\noalign{\smallskip}\hline
\multirow{2}{*}{\ber LONG \eer} & full & 0.0562 & 52.4 & 0.000183 & 0.00269 & 8.37 & 0.0982 \\\cline{2-8}
 & good & 0.0526 & 52.3 & 0.000216 & 0.00314 & 8.22 & 0.0927 \\
\noalign{\smallskip}\hline
\multirow{2}{*}{\ber ELEV \eer} & full & -0.637 & 52.4 & 0.000484 & 0.00269 & 7.63 & -0.42 \\\cline{2-8}
 & good & -0.632 & 52.3 & 0.000588 & 0.00314 & 7.53 & -0.41 \\
\noalign{\smallskip}\hline
\end{tabular}
\end{table}

\begin{table}
\centering
\caption{Linear regression results for $\Upsilon$}
\label{tab:E}       
\begin{tabular}{l|l|l|llllll}
\hline\noalign{\smallskip}
$X_i$ & data & SA & $m$ & $c$ & $\sigma_{mm}$ & $\sigma_{cc}$ & $\sqrt{\langle \chi^2 \rangle}$ & $r_P$ \\
\noalign{\smallskip}\hline\noalign{\smallskip}
\multirow{4}{*}{\ber LAT \eer} & \multirow{2}{*}{\ber full \eer}
   & SMA & 0.441 & 4.52 & 0.0116 & 0.0576 & 6.75 & 0.31 \\\cline{3-9}
 & & SNA & 0.786 & 9.3 & 0.0284 & 0.142 & 14.9 & 0.255 \\\cline{2-9}
 & \multirow{2}{*}{\ber good \eer}
   & SMA & 0.442 & 3.28 & 0.0148 & 0.0723 & 7.35 & 0.282 \\\cline{3-9}
 & & SNA & 0.678 & 6.4 & 0.035 & 0.172 & 15.8 & 0.207 \\
\noalign{\smallskip}\hline
\multirow{4}{*}{\ber LONG \eer} & \multirow{2}{*}{\ber full \eer}
   & SMA & -0.03 & 4.52 & 0.00392 & 0.0576 & 7.08 & -0.0621 \\\cline{3-9}
 & & SNA & 0.0235 & 9.3 & 0.00965 & 0.142 & 15.4 & 0.0224 \\\cline{2-9}
 & \multirow{2}{*}{\ber good \eer}
   & SMA & -0.00781 & 3.28 & 0.00504 & 0.0723 & 7.66 & -0.0146 \\\cline{3-9}
 & & SNA &     0.1 & 6.4 & 0.0119 & 0.172 & 16.1 & 0.0897 \\
\noalign{\smallskip}\hline
\multirow{4}{*}{\ber ELEV \eer} & \multirow{2}{*}{\ber full \eer}
   & SMA & 0.156 & 4.52 & 0.0104 & 0.0576 & 7.04 & 0.122 \\\cline{3-9}
 & & SNA & 0.241 & 9.3 & 0.0256 & 0.142 & 15.3 & 0.0869 \\\cline{2-9}
 & \multirow{2}{*}{\ber good \eer}
   & SMA & 0.142 & 3.28 & 0.0141 & 0.0723 & 7.63 & 0.0952 \\\cline{3-9}
 & & SNA & 0.136 & 6.4 & 0.0329 & 0.172 & 16.1 & 0.0443 \\
\noalign{\smallskip}\hline
\end{tabular}
\end{table}

\clearpage

\clearpage

%
%

\end{document}